\documentclass[journal]{IEEEtran}
\ifCLASSINFOpdf
\else
\fi

\usepackage{cite}
\usepackage{amsmath,amssymb,amsfonts}

\DeclareMathOperator*{\argmin}{arg\,min}
\usepackage{graphicx}
\usepackage{textcomp}
\usepackage{xcolor}
\usepackage[english]{babel}
\usepackage{tabularx}
\usepackage{algorithm}
\usepackage{algpseudocode}
\usepackage{caption}
\usepackage{subcaption}

\newtheorem{assumption}{Assumption}

\newtheorem{remark}{Remark}

\begin{document}
%
\title{{A Finite-State Fixed-Corridor Model for UAS Traffic Management}}
%
%
%

\author{Hamid Emadi,~
        Ella Atkins,
         ~\IEEEmembership{Senior Member,}
        and~Hossein Rastgoftar
\thanks{Hamid Emadi and Hossein Rastgoftar are with the Aerospace and Mechanical Engineering Department, University of Arizona, Tucson, AZ Email: \{hamidemadi, hrastgoftar\}@arizona.edu.}
\thanks{Ella Atkins is with the Aerospace Engineering and Robotics Departments, University of Michigan, Ann Arbor, MI Email: ematkins@umich.edu.}
}

%
%

 \markboth{}
{Shell \MakeLowercase{\textit{et al.}}: Bare Demo of IEEEtran.cls for IEEE Journals}
%



\maketitle


\begin{abstract}
This paper proposes a physics-inspired solution for low altitude Unmanned Aircraft System (UAS) Traffic Management (UTM) in urban areas. We decompose UTM into spatial and temporal planning problems. For the spatial planning problem, we use the principles of Eulerian continuum mechanics to safely and optimally allocate finite airspace to a UAS. To this end, the finite airspace is partitioned into planned and unplanned subspaces with unplanned subspace(s) or zone(s) enclosing buildings and restricted no-fly regions.  The planned subspace is divided into navigable channels that safely wrap unplanned zone(s). We model the airspace planning problem as a Markov Decision Process (MDP)  with states defined based on spatial and temporal airspace features and actions authorizing transitions between safe navigable channels. We apply the proposed traffic management solution to plan safe coordination of small UAS in the airspace above downtown Tucson, Arizona. 
\end{abstract}


%
\IEEEpeerreviewmaketitle

\section{Introduction}
\IEEEPARstart{U}{nmanned} Aerial Systems (UAS) were originally developed for military applications~\cite{newcome2004unmanned}. UAS are also becoming popular in a variety of industrial and academic research applications 
due to benefits such as their agility, low operational cost, and ability to observe and transit through a complex three-dimensional environment. UAS applications include small package delivery~\cite{jung2017analysis}, data acquisition from hazardous environments~\cite{argrow2005uav}, agricultural inspection and chemical application~\cite{tsouros2019review}, aerial surveillance~\cite{semsch2009autonomous}, urban search and rescue~\cite{surmann2019integration}, wildlife monitoring and exploration~\cite{witczuk2018exploring} and urban traffic monitoring~\cite{butilua2022urban}.

To safely integrate UAS into low altitude airspace, the Federal Aviation Administration (FAA) has published rules that restrict or prohibit UAS operators from flying near sensitive regions like airports, stadiums or prisons~\cite{FAA}. 
UAS must also remain clear of manned aircraft airspace corridors, terrain, and infrastructure. A UAS traffic management (UTM) system inspired by manned air traffic management (ATM)~\cite{sridhar1998airspace} has been proposed to manage UAS traffic in low-altitude airspace. UTM has to-date focused on defining a sparse and static set of UAS traffic corridors, but these manually-defined and mapped corridors will significantly limit transiting UAS traffic density and throughput.  Moreover, UTM requires a transition to autonomy and datalink that will no longer support see-and-avoid and voice-based single-UAS traffic coordination.  The UTM framework must therefore include protocols to assure traffic coordination and collision avoidance along with support for high-density, high-throughput operations.  This paper describes a planning strategy to support collision-free UAS transit through high-density airspace.  Related work and a paper overview are provided below.

\subsection{Related Work}


UTM development by the National Aeronautics and Space Administration (NASA) and FAA is summarized in~\cite{prevot2016uas},\cite{rios2016nasa} and~\cite{kopardekar2016unmanned}. In particular, NASA and FAA are developing UTM-specific metrics and protocols to authenticate users, manage datalink and databases, separate UAS from manned aircraft, and provide updated information to users. A candidate UTM concept of operation in ~\cite{jiang2016unmanned} discusses the roles and responsibilities of UTM and UAS pilots. Although the approach in ~\cite{jiang2016unmanned} is fundamentally based on manned air traffic control, relevant methods of UAS control, maneuverability, range, and operational constraints are presented.  Reference ~\cite{sunil2015metropolis} compares airspace safety and capacity with different protocols ranging from free-flight to a network of fixed corridors with pre-planned UAS trajectories meeting separation standards. Issues in deploying UTM for autonomous point-to-point UAS traffic are discussed in~\cite{lundberg2018urban}. 

A primary UTM challenge is to assure operational scalability ~\cite{dao2017evaluation}, particularly for payload transport applications. This has led package transport companies to propose service models of UTM ~\cite{air2015revising}, e.g., low-speed local traffic, high-speed transit layer traffic. Authors in~\cite{battiste2016function} propose a first-come-first-serve procedure to avoid trajectory conflicts. Because UTM will blanket the ground with low-altitude UAS, UTM protocols must also responsibly integrate with a diverse suite of overflown communities mapped by property zoning, a mobile population, navigation signal availability, terrain, man-made infrastructure, and community preferences~\cite{ochoa2022urban}. 


A three-dimensional air corridor system is proposed in~\cite{feng2020uas} to safely manage low-altitude UAS traffic. Reference ~\cite{zhai2021low} presents a computationally efficient  global subdivision method to organize traffic. Four types of low-altitude air routes are designed with a discrete grid transforming the complex spatial computation problem into a spatial database query. Airspace geofencing has been proposed to assure UAS respect no-fly-zones and remain within their allocated airspace volumes \cite{stevens2020geofence, stevens2019geofence, stevens2017specification}. Reference ~\cite{kim2022airspace} presents three-dimensional flight volumization algorithms using computational geometry and offers path planning solutions responsive to dynamic airspace allocation constraints. UAS path planning must be closely coordinated with or performed by UTM to assure  collision-free flight plans compatible with existing traffic flows. Graph search methods such as A* search~\cite{duchovn2014path} and D* ~\cite{stentz1997optimal} can efficiently generate solutions with abstract or local-area search spaces.  Roadmaps such as Voronoi and visibility graphs ~\cite{latombe2012robot} and random sampling approaches ~\cite{burns2005sampling} such as RRT* \cite{noreen2016optimal} reduce search space complexity in 2-D and 3-D environments.  

{\color{black}Markov decision process (MDP) is a discrete system decision making framework applicable to situations where outcomes are not deterministic. The MDP has been used in a variety of applications including but not limited to finance, maintenance, queue management and robotics. MDP models can be used for path planning given a finite discrete state-space. The authors in~\cite{temizer2010collision} and~\cite{jeannin2017formally} have proposed a collision avoidance system using an MDP model. Reference ~\cite{ragi2013uav} proposes an MDP for UAS path planning to track multiple ground targets in a dynamic environment.

Researchers have developed different methods for UAS coordination. For example, containment control~\cite{li2008formation}, consensus-based control~\cite{li2018nonlinear}, partial differential-based approach~\cite{kim2008pde}, graph-based methods~\cite{wang2018distance} and continuum deformation approach~\cite{rastgoftar2017continuum}\cite{rastgoftar2019physics}. We adopt a compact Eulerian continuum mechanics model for UAS coordination in this paper. }





\subsection{Contributions and Outline}
This paper proposes defining UAS traffic coordination for UTM as a spatiotemporal planning problem. For spatial planning, we define UAS coordination as an ideal fluid flow governed by the Laplace partial differential equation (PDE) with inspiration from ~\cite{rastgoftar2019physics}. Terrain, buildings, and infrastructure are wrapped by airspace obstacles (i.e., no-fly zones) through which we propose the design of fixed airway corridors.  In particular, we divide the airspace into different layers and assign each UAS to transit in a fixed altitude layer along that layer's prescribed traffic flow streamline. Transitions between air corridor layers are permitted at a cost that encourages each UAS to remain in a single layer when possible. For temporal planning, we define an MDP to authorize safe UAS transitions between air corridors in a centralized manner consistent with current concepts proposed for community-based UTM. Compared to the existing literature and the authors' previous work, this paper offers the following novel contributions: 
\begin{enumerate}
    \item We propose a UTM architecture that includes time-invariant air corridor layers for transiting UAS traffic.  Specifically, obstacle-free air corridor geometries are defined by solving a Laplace PDE that safely wraps buildings and no-flight-zones at low computation cost,
    \item {\color{black} We propose an MDP-based collision-free multi-vehicle path planning strategy that applies a first-come-first-serve prioritization to UAS airway corridor allocation. }
\end{enumerate}


This paper is organized as follows. 
Section~\ref{sec: Problem Statement} provides a problem statement followed by a description of our methodology in Section~\ref{sec: Methodology}. Operation of the proposed layered UTM airspace is summarized in Section \ref{operation}.   Simulation results are presented in Section~\ref{sec: Simulation} followed by a brief conclusion in Section~\ref{sec: Conclusion}.


\section{Problem Statement}\label{sec: Problem Statement}
This paper develops a physics-inspired UTM solution to maximize safe low-altitude airspace occupancy by small UAS. Our proposed solution 
defines  UAS routing in UTM as a spatiotemporal planning problem. For spatial planning,  UAS coordination is defined by an ideal fluid flow pattern with potential and stream functions obtained by solving  Laplace PDEs \cite{rastgoftar2019physics, rastgoftar2020fault}. This solution offers the following advantages:
\begin{enumerate}
    \item The streamlines  define the boundaries of air corridors that safely wrap building and no-fly-zones in low-altitude airspace.
    \item{The system can be solved in real-time to guarantee collision avoidance given UAS failures and dynamic evolution of local airspace no-fly zone geometry.}
\end{enumerate}

For temporal planning, we apply an MDP formulation to manage UAS coordination by optimally allocating air corridors to UAS in a first-come-first-serve prioritization. 
This work makes the following simplifying assumption.
\begin{assumption}
Airspace corridor design and allocation is centralized. Each UAS is connected to single local UTM cloud computing system managing low-altitude airspace for that region.
\end{assumption}





\section{Methodology}\label{sec: Methodology}
This section presents a mathematical framework for UAS path planning for different tasks in a 3-D obstacle-laden environment. To this end, we first define fixed air corridors by treating UAS coordination as ideal fluid flow that safely wraps unplanned airspace in Section  \ref{Time Invariant Navigable Channels}. 
Then, we define an MDP  to optimally allocate air corridors to the UAS requesting passage through the managed airspace volume.   

\subsection{Spatial Planning: Air Corridor Generation Using Fluid  Flow  Navigation}\label{Time Invariant Navigable Channels}
Section \ref{Ideal Fluid Flow Pattern:} presents the foundations of ideal fluid flow coordination. Next, Section \ref{Air Corridor Generation} discusses how fluid flow coordination can be applied to generate safe air corridors in urban low-altitude airspace.

\subsubsection{Ideal Fluid Flow Pattern}\label{Ideal Fluid Flow Pattern:}

\begin{figure}[ht]
    \centering
    \includegraphics[width=0.45\textwidth]{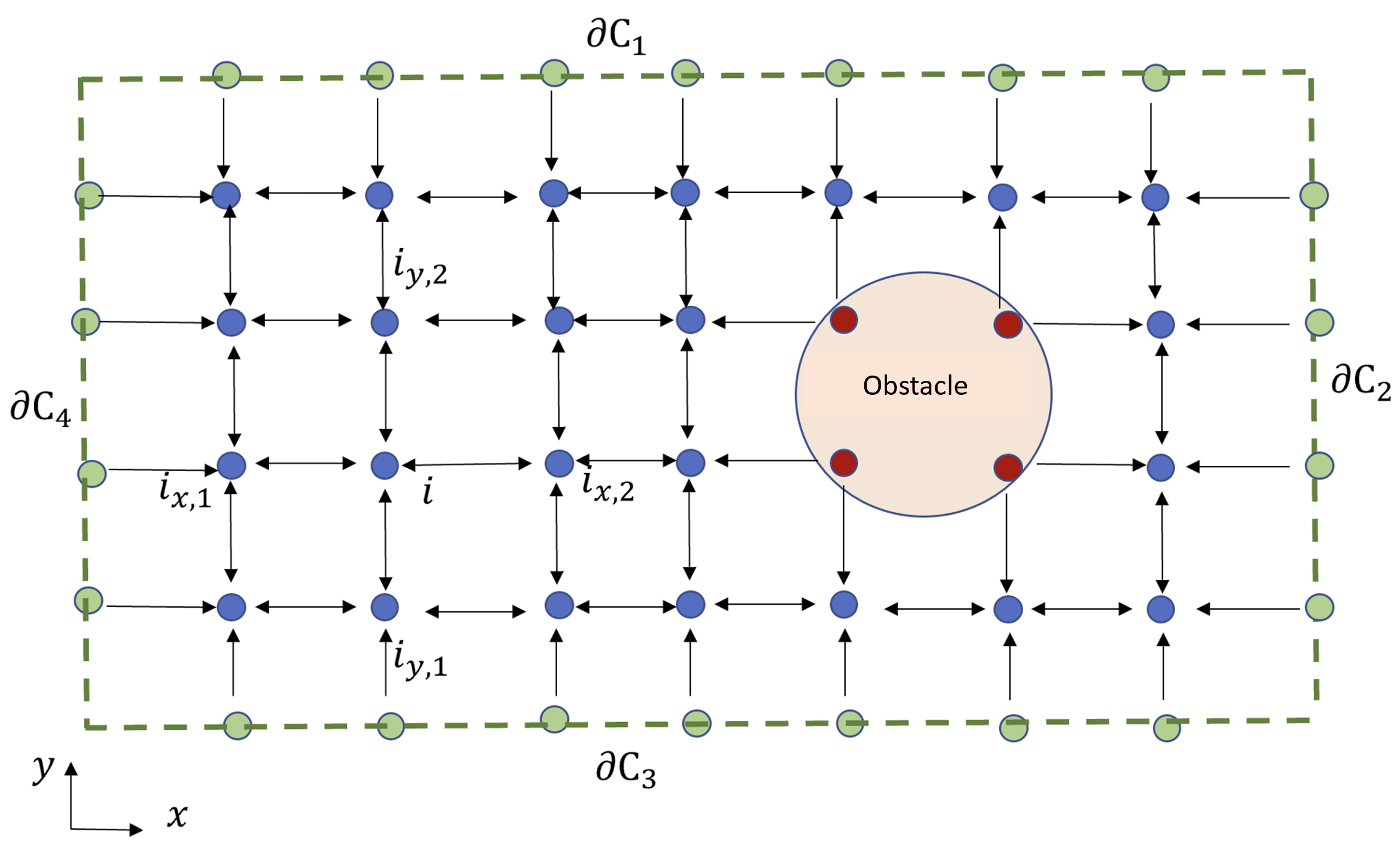}
    \caption{Directed graph $\mathcal{G}$ discretizes the $x-y$ plane. Green, blue and red nodes correspond respectively to the boundary, feasible space and single circular obstacle.}
    \label{fig:grid}
\end{figure}

In this work, we treat UAS as particles in an ideal flow moving on streamlines that wrap unplanned airspace zones. Here, the unplanned zones represent buildings and restricted flight areas \cite{rastgoftar2019physics}.
Ideal fluid flow is defined over compact set $\mathcal{C}\subset \mathbb{R}^2$, where $\mathcal{C}$ is a projection of a finite airspace on a $2$-D plane. 
Without loss of generality, we assume that $\mathcal{C}\subset \mathbb{R}^2$ lies in the $x-y$ plane (see Fig. \ref{fig:grid}). Assuming the airspace contains $n_o$ unplanned zones, their projections on $\mathcal{C}$ are defined by disjoint closed sets of $\mathcal{O}_1,\dots,\mathcal{O}_{n_o}$. Let complex variable $\mathbf{z}=x+\mathbf{i}y$  denote position in the $x-y$ plane. We obtain potential function ${\Phi}\left(x,y\right)$ and stream function ${\Psi}\left(x,y\right)$ of the ideal fluid flow field by defining a conformal mapping 
\begin{eqnarray}
f\left(\mathbf{z}\right)={\Phi}\left(x,y\right)+\mathbf{i} {\Psi}\left(x,y\right)
\end{eqnarray}
with $\Phi(x,y)$ and $\Psi(x,y)$ that satisfy the Laplace PDE and Cauchy-Riemann conditions:
\begin{eqnarray}\label{Laplac}
    \nabla^2 \Psi = 0, \quad \nabla^2 \Phi = 0
\end{eqnarray}
\begin{eqnarray}\label{Cauchy-Riemann}
     \frac{\partial\Phi}{\partial x} = \frac{\partial\Psi}{\partial y}, \quad \frac{\partial\Phi}{\partial y} = -\frac{\partial\Psi}{\partial x}
\end{eqnarray}

Using the ideal fluid flow model~\cite{rastgoftar2019physics}, $x$ and $y$ components of the $i^\text{th}$ UAS are constrained to slide along stream curve $\Psi_i$ defined as follows:
 \begin{equation}\label{eq: psi initial}
     \Psi_{i} \triangleq \Psi = \Psi(x_i(t_0),y_i(t_0))
 \end{equation}
where $x_i(t_0)$ and $y_i(t_0)$ are $x$ and $y$ components of the $i^\text{th}$ UAS position at reference time $t_0$ when it enters $\mathcal{C}$ through a boundary point.  
We can use analytic and numerical approaches to define $\Phi(x,y)$ and $\Psi(x,y)$ over $\mathcal{C}$ as described next.


\paragraph{Analytic Solution}

Unplanned or "no-fly" airspace zones defined by $\mathcal{O}_1,\dots,\mathcal{O}_{n_o}$ can be safely wrapped by defining $\Phi$ and $\Psi$ as the real and imaginary parts of complex function
\begin{equation}
    f\left(\mathbf{z}\right)=\sum_{i=1}^{n_o}\left(\mathbf{z}-\mathbf{z}_i+\dfrac{r_i^2}{\mathbf{z}-\mathbf{z}_i}\right),
    \label{eq: complex function}
\end{equation}
where $\mathbf{z}_i=x_i+\mathbf{j}y_i$ and $r_i>0$ denote the nominal position and size of the  $i$-th unplanned zone, respectively. Here, $r_i$ must be sufficiently large so that the $i$-th obstacle is safely enclosed. 

\begin{remark}
For $n_o=1$, a single compact unplanned region existing in $\mathcal{C}$ is wrapped by a circle of radius $r_i$ with center positioned at $\mathbf{z}_i$. {\color{black}However, when $n_o>1$, unplanned zones are not wrapped with exactly a  circular area. Therefore, analytic solution~\eqref{eq: complex function} cannot be used for environments containing arbitrary obstacles. }
\end{remark}


\paragraph{Numerical Solution}
When environments contains obstacles with arbitrary non-circular sections, we use the finite difference approach to determine $\Phi$ and $\Psi$ values over the motion space. The finite-difference method discretizes the governing PDE and the environment by replacing the partial derivatives with their approximations. We therefore uniformly discretize $\mathcal{C}$ into small regions with increments in the $x$ and $y$ directions given as $\Delta x$ and $\Delta y$, respectively. We use graph $\mathcal{G}\left(\mathcal{V},\mathcal{E}\right)$ to uniformly discretize  $\mathcal{C}$ where $\mathcal{V} =\{1,\dots,m\}$ and $\mathcal{E} \subseteq \mathcal{V} \times \mathcal{V}$ define nodes and edges of $\mathcal{G}$, respectively. 

We express set $\mathcal{V}=\left\{1,\cdots,m\right\}$ as $\mathcal{V}=\mathcal{V}_B\bigcup\mathcal{V}_I\bigcup \mathcal{V}_O$ where disjoint sets $\mathcal{V}_B=\{1,\dots,{m_B}\}$, $\mathcal{V}_I= \{{m_B+1},\dots,{m_B+m_I}\}$, and $\mathcal{V}_O=\{{m_B+m_I+1},\dots,{m}\}$ identify \textit{boundary}, \textit{interior}, and \textit{obstacle} nodes, respectively (i.e. $m=\left|\mathcal{V}\right|$ is the total number of nodes used for discretization of $\mathcal{C}$).
Fig. \ref{fig:grid} shows a uniform discretization of rectangular domain $\mathcal{C}$ with the boundaries denoted by  $\partial\mathcal{C}_1$, $\partial\mathcal{C}_3$, $\partial\mathcal{C}_2$, and $\partial\mathcal{C}_4$. Assuming the UAS objective is to safely move from left to right, we impose the following conditions (constraints) on $\Psi$ over $\mathcal{V}_B$ and $\mathcal{V}_O$:
\begin{eqnarray} \label{boundary conditions}
    \Psi(j) =  
    \left\{\begin{matrix}
       K_1y_j+K_2 & j\in \partial\mathcal{C}_2 \bigcup \partial\mathcal{C}_4 \\ 
       K_3x_j+K_4 & j\in \partial\mathcal{C}_1 \bigcup \partial\mathcal{C}_3 \\ 
        0 &  j\in \mathcal{V}_O
    \end{matrix}\right.
\end{eqnarray}
Here, $y_j$ is the $y$ component of node $j$ position{\color{black}; $K_1$, $K_2$, $K_3$, and $K_4$ are constant parameters that are assigned so that the streamlines are directed along the $x$ or $y$ axis.  When streamlines are directed along the $x$ axis, $K_3=0$ and $K_1\neq 0$. Also, $K_1=0$ and $K_3\neq 0$ when the streamlines are directed along the $y$ axis.}  From the above expression, $\Psi$ is constant over $\partial\mathcal{C}_1$ and  $\partial\mathcal{C}_2$ which in turn implies that $\partial\mathcal{C}_1$ and $\partial\mathcal{C}_2$ are the boundary streamlines.

By substituting the approximated derivatives from the Taylor series to~\eqref{Laplac}, stream value function $\Psi_i$ at node $i\in \mathcal{V}_I$ satisfies the following equation:
\begin{eqnarray}\label{discrete_laplace_phi}
    \frac{\Psi_{i_{x,1}}-2\Psi_i+\Psi_{i_{x,2}}}{{\Delta x}^2} + \frac{\Psi_{i_{y,1}}-2\Psi_i+\Psi_{i_{y,2}}}{{\Delta y}^2} = 0,
\end{eqnarray}
where $\Psi_{i_{x,1}}$ and $\Psi_{i_{x,2}}$ are $\Psi$ values at the neighbor nodes in the $x$ direction, i.e. $\left(i_{x,1},i\right),\left(i_{x,2},i\right)\in \mathcal{E}$. Similarly, $\Psi_{i_{y,1}}$ and $\Psi_{i_{y,2}}$ are the $\Psi$ values at the neighbor nodes in the $y$ direction, i.e. $\left(i_{y,1},i\right),\left(i_{y,2},i\right)\in \mathcal{E}$.

Let $\boldsymbol{\Psi} = \begin{bmatrix}
\Psi_1,\dots,\Psi_m
\end{bmatrix}^T$ represent the nodal vector of the potential function. Equation \eqref{discrete_laplace_phi} can then be written in compact form 
\begin{eqnarray}\label{eq: L * phi=0}
    {\mathbf{L}}\boldsymbol{\Psi} = \boldsymbol{0}
\end{eqnarray}
where ${\mathbf{L}=\left[L_{ij}\right]}\in \mathbb{R}^{m\times m}$ is the Laplacian matrix of graph  $\mathcal{G}$ with $\left(i,j\right)$ entry 
\begin{eqnarray}
    L_{ij}=\left\{\begin{matrix}
     \text{deg}(i)&  i=j  \\ 
     -1&  i\neq j,(i,j)\in \mathcal{E}\\
     0&  \text{otherwise}
\end{matrix}\right.
\end{eqnarray}
where $\text{deg}(i)$ is the in-degree of node $i$. According to~\cite{veerman2020primer} the multiplicity of eigenvalue 0 of ${\mathbf{L}}$ is equal to the number of maximal reachable vertex sets. In other words, multiplicity of zero eigenvalues is the number of trees needed to cover graph  $\mathcal{G}$. Therefore, matrix ${\mathbf{L}}$ has $m_B+m_O$ eigenvalues equal to 0. Hence, the rank of $L$ is $m-m_B+m_O$. 

\begin{figure}[ht]
    \centering
    \includegraphics[width=0.45\textwidth]{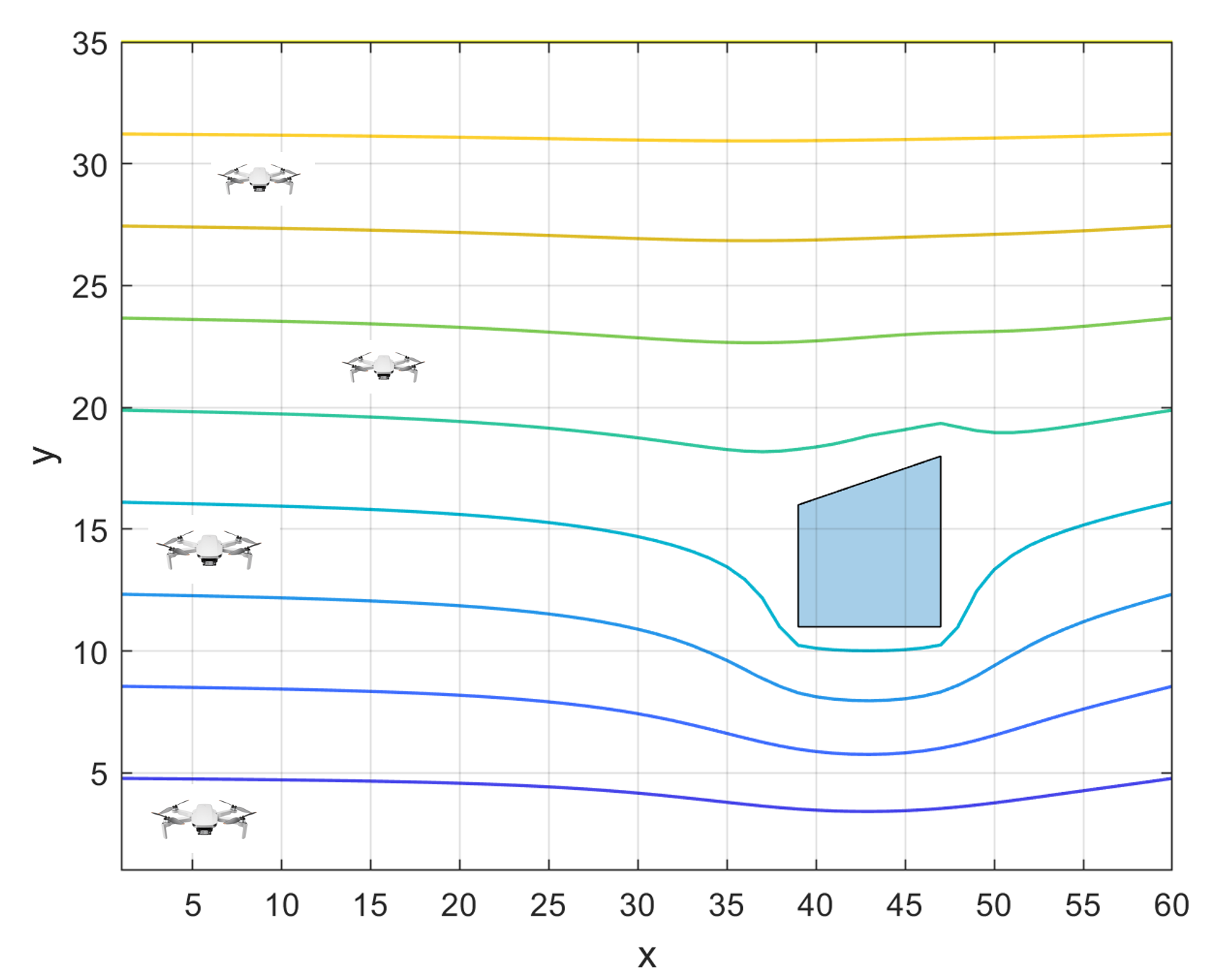}
    \caption{Streamlines in the $x-y$ plane for an environment with a polygonal obstacle.}
    \label{fig: streamline}
\end{figure}

{\color{black}Let $\Bar{\mathbf{\Psi}} = \left[\psi_{m_B+1},\dots,\psi_{m_B+m_I}\right]$ denote the vector of $\psi$ values corresponding to the interior nodes. Since rank $L$ is $m-m_B+m_O$, \eqref{eq: L * phi=0} can be solved for $\Bar{\mathbf{\Psi}}$. Details of this numerical approach are presented in~\cite{rastgoftar2019physics}.}  Fig.~\ref{fig: streamline} shows the streamlines in a rectangular environment wrapping a polygonal obstacle  obtained with the numerical approach presented above. 

\subsubsection{Air Corridor Generation}\label{Air Corridor Generation}

We decompose the 3-D environment into $n_l$ layers, identified by $\mathcal{C}_1,\dots,\mathcal{C}_{n_l}$, corresponding to different altitudes. Mathematically speaking, $\mathcal{C}_i\subset \mathbb{R}^2$ is a horizontal floor parallel to the $x-y$ plane at altitude $z=h_i$.

Let $\mathcal{O}^j_i\subset \mathcal{C}_i$ be the projection of  unplanned zone $\mathcal{O}_j$ on $\mathcal{C}_i$. Using the numerical approach expressed in Section~\ref{Time Invariant Navigable Channels}, we can safely exclude $\mathcal{O}^1_i\bigcup \cdots \mathcal{O}^{n_o}_i$ by obtaining stream function  $\Psi_i(x,y)$ over $\mathcal{C}_i$, and discretize the planned space 
\[
\mathcal{P}_i=\mathcal{C}_i\setminus \left(\mathcal{O}^1_i\bigcup \cdots \mathcal{O}^{n_o}_i \right),\qquad \forall i\in \{1,\dots,n_l\}
\]
into a finite number of corridors with the boundaries obtained by level curves with $\Psi_i(x,y)=\mathrm{constant}$. 

\subsection{Temporal Planning: Optimal Allocation of Air Corridors to UAS}\label{Temporal Planning: Optimal Allocation of Air Corridors to UAS}
We define an MDP  to maximize the usability of the low-altitude airspace through optimal allocations of air corridors to UAS.  Note that although the below formulation supports a general stochastic dynamic programming (SDP) or MDP model, we later define case studies relying on a
dynamic programming model for which the MDP/SDP state transition function is deterministic rather than stochastic. 

The generalized MDP models decision-making in discrete environments with stochastic or partially stochastic outcomes and is defined by tuple 
\[
\mathcal{M}=\{\mathcal{S},\mathcal{A}, P, J,\gamma\}
\]
with the following elements: 
\begin{enumerate}
    \item Finite set of states $\mathcal{S}$;
     \item Finite set of actions $\mathcal{A}$;
     \item State transition dynamics defined by probability tensor
     \[
     P=\bigcup_{s,s'\in \mathcal{S},a\in \mathcal{A}}P_a(s,s')
     \]
     where $P_a(s,s')$ assigns transition from current state $s\in \mathcal{S}$ to state $s'\in \mathcal{S}$ under action $a\in \mathcal{A}$;
     \item Cost function
     \[
     J=\bigcup_{s,s'\in \mathcal{S},a\in \mathcal{A}}J_a(s,s')
     \]
     where $J_a(s,s')$ assigns numerical cost at state $s\in \mathcal{S}$ under action $a\in \mathcal{A}$;
     \item Discount factor $\gamma\in \left[0,1\right]$.
\end{enumerate}
Note that cost function $J$ can equivalently be defined by a negative reward function $R$ in an MDP. {\color{black} The MDP policy specifying a mapping from states to actions maximizes expected value at each state. Note that some researchers \cite{agarwal2021markov} define an MDP formulation that minimizes cost and expected cost instead.  We use a minimum cost MDP formulation in this paper.}  
MDP value function $V:\mathcal{S}\rightarrow \mathbb{R}^+$ defines total {\color{black} expected value} corresponding to the sequence of states ${\color{black}s}=(s_1,\dots, s_t,\dots)$ and actions $a=(a_1,\dots,a_t,\dots)$:
\begin{eqnarray}
V = \sum_{t=0}^{\infty}\gamma^t J_{a_t}(s_t,s_{t+1}).
\end{eqnarray}
Using the value iteration algorithm, we can compute a function $V_i(s): \mathcal{S}\rightarrow \mathbb{R}^+$ that associates to each state $s$ a lower bound to the optimal cost $V^*(s)$. In particular, by updating $V_i(s)$ in the following way 
\begin{eqnarray}
V_{i+1}(s) = \min_{a}\sum_{s'\in \mathcal{S}}{P_a(s,s')(J_a(s,s')+ \gamma V_i(s') )}
\end{eqnarray}
$V_{i}(s)$ converges monotonically and in polynomial time to $V^*(s)$. Threshold $\epsilon$ specifies the numerical convergence requirement for value in each state:
\begin{eqnarray}
    V^* \approx \min\limits_{s\in \mathcal{S}} |V_{i+1}(s)-V_i(s)| \leq \epsilon.
\end{eqnarray}
The optimal policy $\pi^*(s)$ is defined as the sequence of actions that provide the optimal total cost $V^*(s)$ starting at state $s$ and is computed from: 
\begin{eqnarray}\label{optimalpolicy}
\pi^*(s) = \argmin\limits_{a\in \mathcal{A}}\sum_{s'\in \mathcal{S}}{P_a(s,s')(J_a(s,s')+ \gamma V^*(s') )}.
\end{eqnarray}



\begin{assumption}\label{assum2}
In this paper, we assume that $P_a(s,s')\in \left\{0,1\right\}$ for every $a\in \mathcal{A}$ and $s,s'\in \mathcal{S}$. Therefore, transition over the state space $\mathcal{S}$ is deterministic under each action $a\in \mathcal{A}$.
\end{assumption}
Although, we assume that transitions over the state space are deterministic, we can indirectly incorporate uncertainty into planning by updating the geometry of the unplanned space without changing the dimension of the state space. In other words, if a UAS cannot admit or follow the desired corridor assigned by the authorized decision-maker, it is contained by an unplanned airspace and safely excluded from the planned airspace.  Note that this problem has been previously investigated by the second and third authors in Ref. \cite{rastgoftar2019physics}.

For the UAS traffic management, without loss of generality, the low altitude airspace is projected on eight layers ($n_l=8$), denoted by $\mathcal{C}_1$, $\cdots$, $\mathcal{C}_8$, where:
\begin{enumerate}
\item Streamlines are elongated along the $x$ axis on $\mathcal{C}_1$, $\mathcal{C}_3$, $\mathcal{C}_5$, and $\mathcal{C}_7$;
\item Streamlines are elongated along the $y$ axis on $\mathcal{C}_2$, $\mathcal{C}_4$, $\mathcal{C}_6$, and $\mathcal{C}_8$.
\end{enumerate}
We define $\mathcal{L}_i$ as a finite set identifying the air corridors at $\mathcal{C}_i\subset \mathbb{R}^2$ ($i\in \left\{1,\cdots,8\right\}$). Also, we define $\mathcal{X}$ and $\mathcal{Y}$ as finite sets representing discrete values of $x$ and $y$ coordinates, respectively, and finite set $\mathcal{T}$ defines future discrete times. 

The state set $\mathcal{S}$ is finite and defined by
\[
\mathcal{S}=\left(\left(\bigcup_{i=1,3,5,7}\left(\mathcal{L}_i\times \mathcal{X}\right)\right)\bigcup \left(\bigcup_{i=2,4,6,8}\left(\mathcal{L}_i\times \mathcal{Y}\right)\right)\right)\times \mathcal{T}.
\]
where ``$\times$'' is the Cartesian product symbol.

We define four possible actions $\mathcal{A}=\{a_1,a_2,a_3,a_4\}$ with the following functionality: 
\begin{itemize}
    \item $a_1$: Move forward in the current corridor,
    \item $a_2$: Stay at the current position for the next time,
    \item $a_3$: Move to the next higher level, 
    \item $a_4$: Move to the next lower level. 
\end{itemize}
Note that the actions are constrained to satisfy the following limitations:
\begin{enumerate}
    \item At the highest level $a_3$ must \underline{not} be selected.
    \item At the lowest level $a_4$ must \underline{not} be selected.
    \item Transition from the current state $s\in \mathcal{S}$ to the next state $s'\in \mathcal{S}$ is allowed only if $s'$ has not already been allocated to another UAS.
\end{enumerate}
Without loss of generality, case studies in this paper assume transitions over the states are deterministic, which in turn implies that probability $P_a\left(s,s'\right)$ is a binary variable, either $0$ or $1$ for $a\in \mathcal{A}$ and $s,s'\in \mathcal{S}$.
More specifically, when transition from $s$ to $s'$ is feasible, $P_a(s,s')=1$, otherwise $P_a(s,s')=0$. We assume that the next state $s'$ is feasible, if $d(s,s')\leq \delta_0$, where $\delta_0$ is a threshold value, and $d(s,s')$ is the Euclidean distance between two states $s$ and $s'$ that is defined as follows:
\[
d\left(s,s'\right)=\sqrt{\left(x-x'\right)^2+\left(y-y'\right)^2},
\]
where $\left(x,y\right)$ and $\left(x',y'\right)$ are positions associated with current state $s\in \mathcal{S}$ and next state $s'\in \mathcal{S}$, respectively.
{\color{blue}

}
To optimally allocate air corridors to a new UAS, we define cost $ J_a(s,s')$ as follows:
\begin{eqnarray}\label{eq: cost function}
    J_a(s,s') = d(s',s_g) + \alpha_a J_0,\qquad a\in \mathcal{A}
\end{eqnarray}
where $s_g$ is the target state for a new UAS, $d(s',s_g)$ is the metric distance between the next state $s'$ and target state $s_g$, and $J_0$ is constant and considered to penalize unnecessary layer change, where $\alpha_a$ is a binary variable defined as follows:
\begin{equation}
    \alpha_a=\begin{cases}
    1&a=a_3,a_4\\
    0&a=a_1,a_2\\
    \end{cases}
    ,\qquad a\in \mathcal{A}.
\end{equation}
In this paper, we choose $\gamma=1$ to optimally assign the air corridors to the UAS. The optimal policy $\pi^*(s)$ obtained by  \eqref{optimalpolicy} is assigned by the value iteration method.
\begin{figure*}[ht]
    \centering
    \includegraphics[width=1\textwidth]{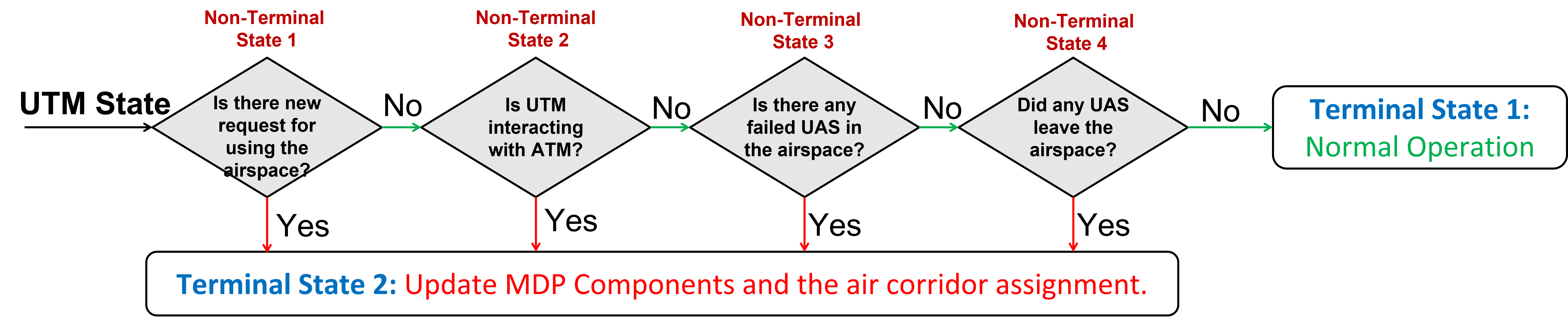}
    \caption{The state machine used to manage MDP updates for safe allocation of UAS to streamline-based airspace corridors.}
    \label{state,achine}
\end{figure*}

\begin{figure}
     \centering
     \begin{subfigure}[b]{0.5\textwidth}
         \centering
         \includegraphics[width=\textwidth]{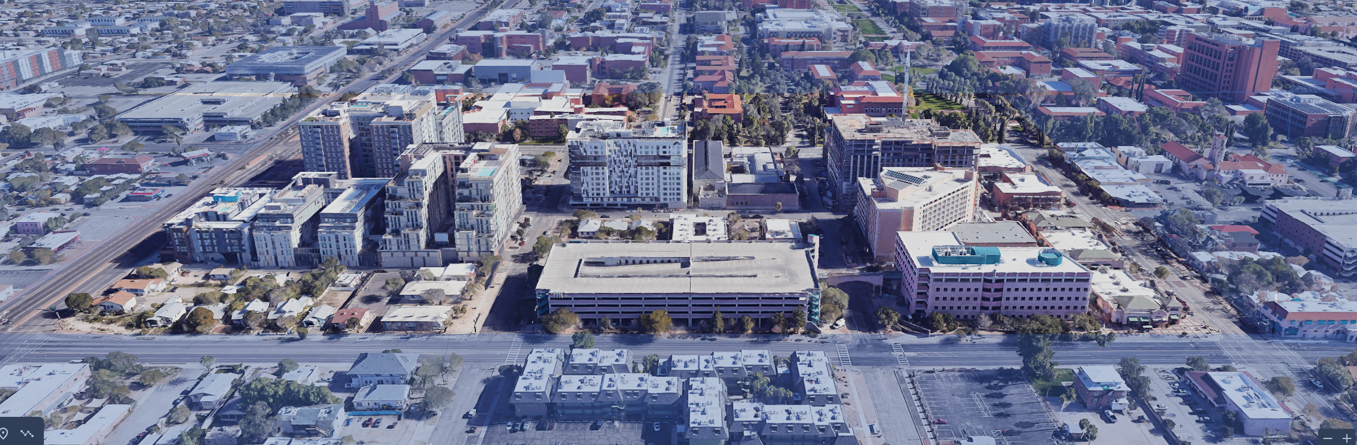}
         \caption{}
         \label{fig: real map1}
     \end{subfigure}
     \hfill
     \begin{subfigure}[b]{0.5\textwidth}
         \centering
         \includegraphics[width=\textwidth]{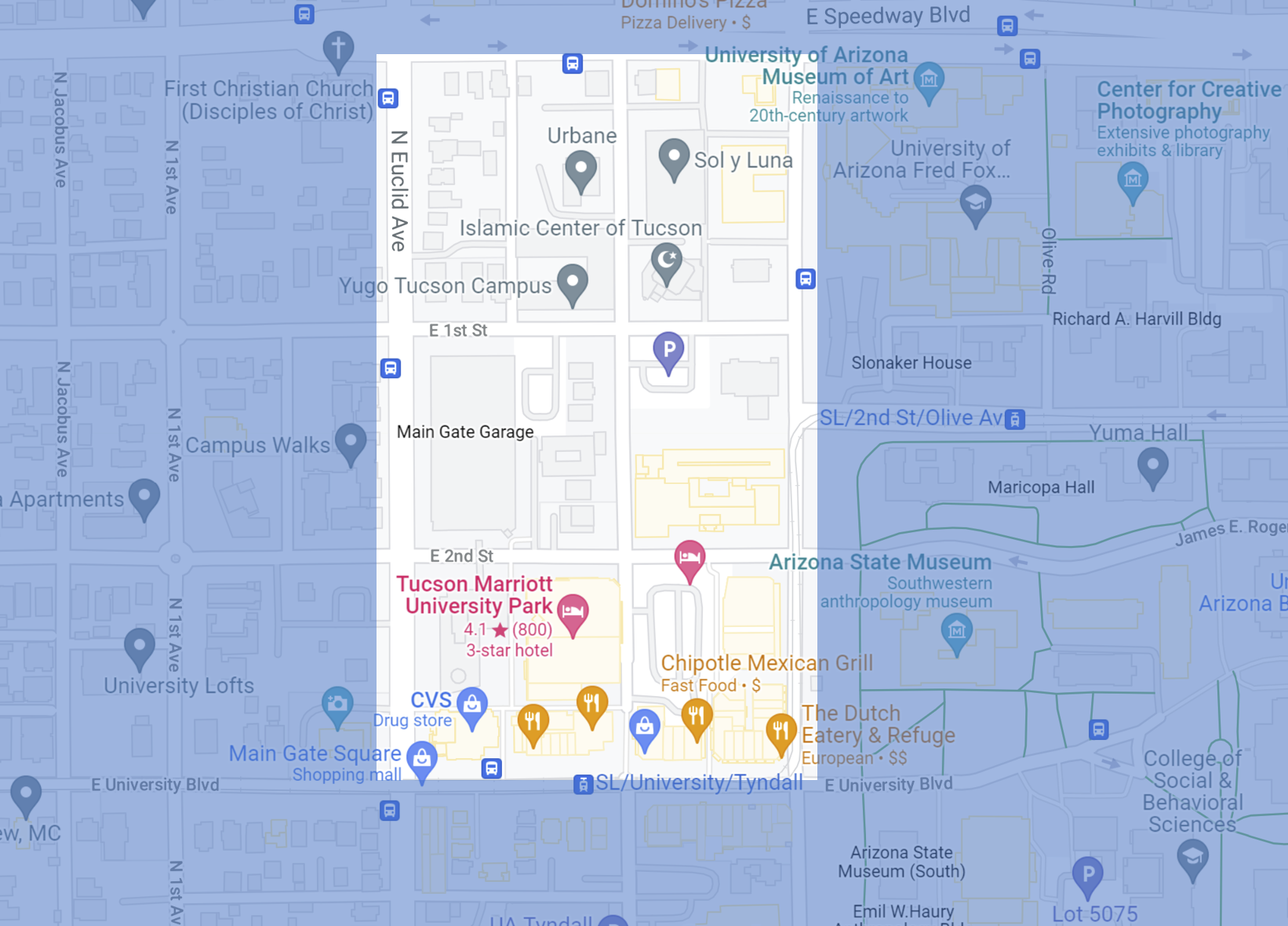}
         \caption{}
         \label{fig: 2D map}
     \end{subfigure}
        \caption{Modeled urban environment in central Tucson-Arizona.  (a) Image from Google Earth. (b) 2-D map from Google Maps.}
        \label{fig: real map}
\end{figure}

\begin{figure}
     \centering
     \begin{subfigure}[b]{0.5\textwidth}
         \centering
         \includegraphics[width=\textwidth]{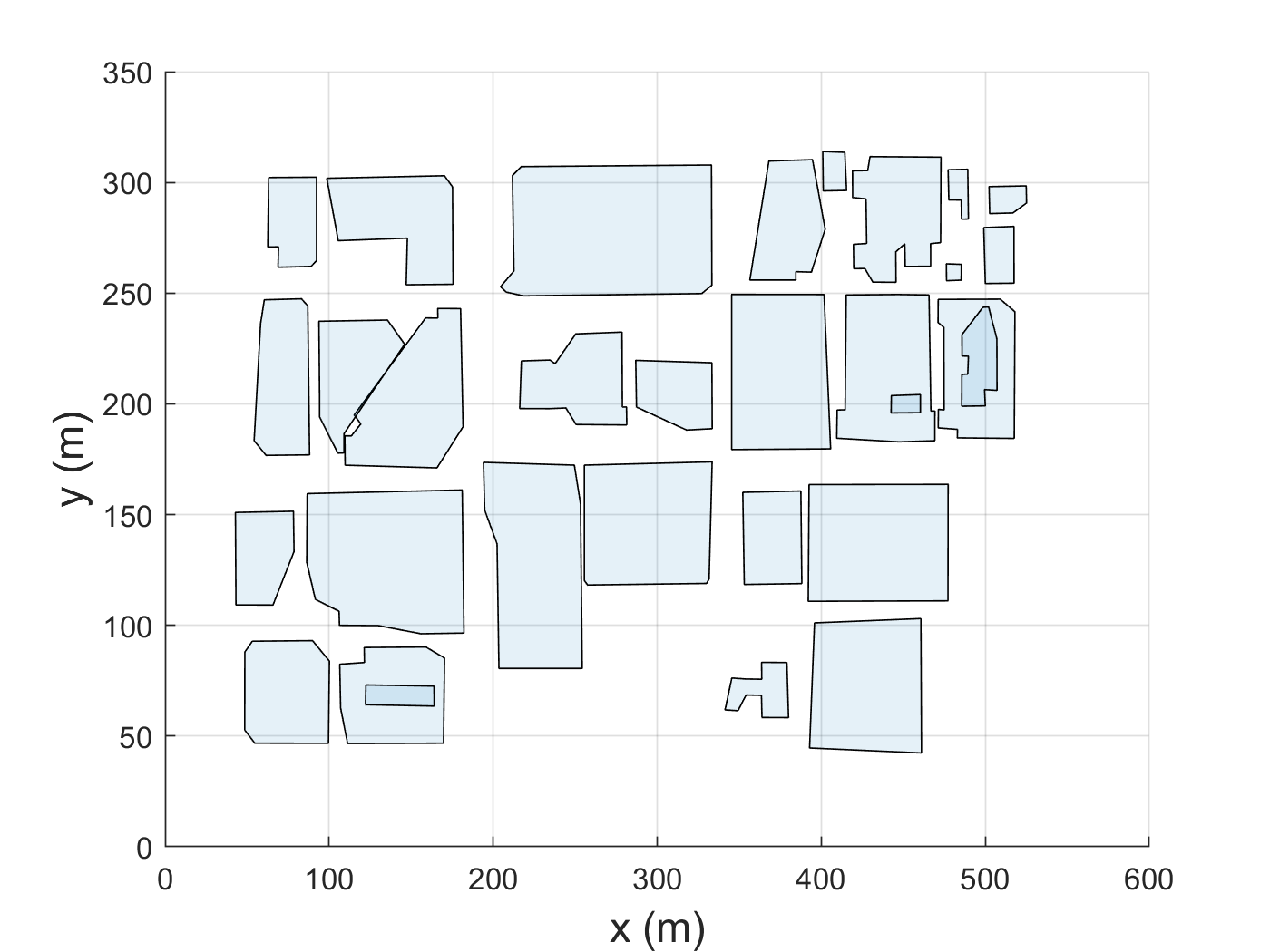}
         \caption{}
         \label{fig: 2D map floor level}
     \end{subfigure}
     \hfill
     \begin{subfigure}[b]{0.5\textwidth}
         \centering
         \includegraphics[width=\textwidth]{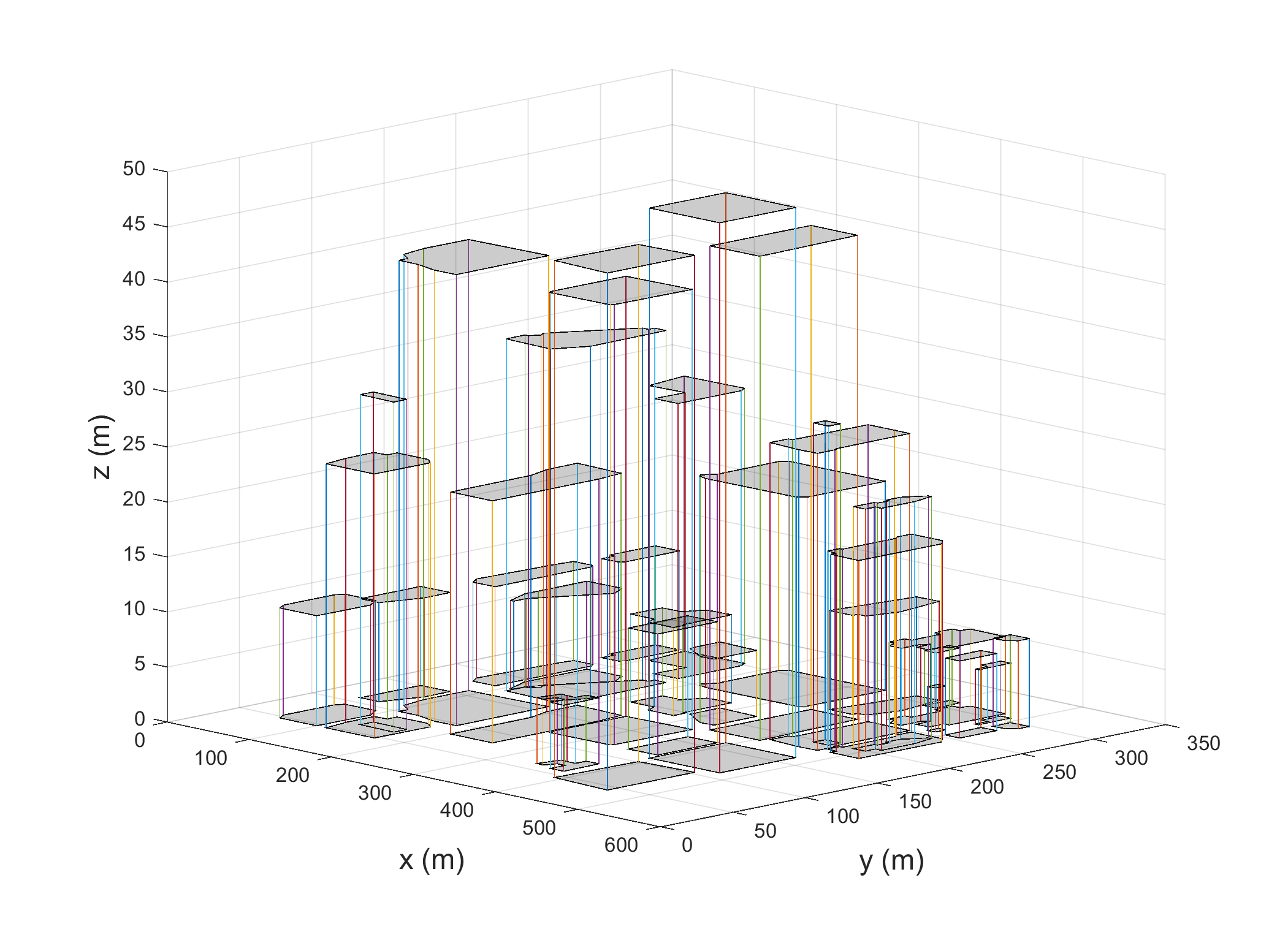}
         \caption{}
         \label{fig: 3D matlab}
     \end{subfigure}
        \caption{(a) Cross section of the floor level of the modeled environment in $x-y$ plane. (b) 3-D model of the environment in MATLAB.}
        \label{fig: real map}
\end{figure}
\section{UTM Operation}\label{operation}
To safely allocate the airspace to the UAS requesting airspace access, we  prioritize the airspace usability by the existing UAS and apply a first-come-first-serve strategy to authorize access for the new UAS. Air corridors can be optimally allocated to UAS using the MDP approach presented in Section  \ref{Temporal Planning: Optimal Allocation of Air Corridors to UAS}. Computational cost is reasonable for real-time policy updates because in most air corridors are already assigned to existing UAS inaccessible by updating the MDP transitions when there is new request for using the airspace. Therefore, the proposed MDP approach assigns airways only to a single UAS after the request is submitted. We apply the state machine shown in Fig.~\ref{state,achine} to safely and resiliently implement our proposed UTM system. This state machine consists of two terminal states and four non-terminal states with definitions given in Table~\ref{tab:my_label}. 

Algorithm \ref{alg: MDP} presents the functionality of our proposed physics-inspired UTM system. If no non-terminal (NT) state is satisfied, the current policy $\pi^*(s)$ for air corridor allocation is acceptable. If non-terminal (NT) state $2$ or NT state $3$ is satisfied, we perform the following steps:
\begin{itemize}
    \item Consider a failed UAS or no-fly zone temporarily allocated to ATM as temporary unplanned airspace, and revise definitions of the air corridors assigned by $\mathcal{L}_1, \cdots, \mathcal{L}_8$.
    \item Update definitions of the state set $\mathcal{S}$, transition probabilities, and  cost.
    \item Update the optimal policy $\pi^*(s)$ by solving Eq. \eqref{optimalpolicy}.
\end{itemize}

If NT state $1$ or NT state $4$ is satisfied, we do not need to revise the state set $\mathcal{S}$ and action set $\mathcal{A}$. However,  cost and transition functions will change and the updated policy is obtained  by solving \eqref{optimalpolicy}.





\begin{table*}[]
    \centering
    \caption{Terminal and non-terminal states of the state machine used for to manage optimal UAS air corridor allocation.}
    \begin{tabular}{|c|c|}
    \hline
     & Implication  \\
    \hline
        Terminal State 1 & Normal Operation: Current optimal allocation of air corridors
are acceptable. 
 \\
\hline
        Terminal State 2 & Update definitions of states, actions, transition probabilities, and cost; update the air corridor assignment. 
 \\
 \hline
       Non-Terminal State 1 & Check if the UTM interfaces with ATM.

 \\
  \hline
       Non-Terminal State 2 & Check if there is a new request for entering or departing the airspace.
 \\
  \hline
       Non-Terminal State 3 & Check if  there are any failed UAS in the airspace.

 \\
  \hline
       Non-Terminal State 4 & Check if  there is a new request for entering or departing the airspace.
 \\
 \hline
    \end{tabular}
    \label{tab:my_label}
\end{table*}

\begin{algorithm}
\caption{Physics-Inspired UTM System}\label{alg: MDP}
\begin{algorithmic}
    \State \textbf{Input} {Current UTM state}
    \State \textbf{Input} {Unplanned airspace $\mathcal{O}_1$ through $\mathcal{O}_{n_o}$}
    \State Specify air corridors defined by $\mathcal{L}_1$ through $\mathcal{L}_8$.
    \State Define MDP components.
    \State \textbf{Output} Assign $\pi^*(s)$ for every $s\in \mathcal{S}$ using Eq. \eqref{optimalpolicy}.
    \If{{\underline{No}} non-terminal state is satisfied}
    \State Normal operation
    \Else
        \If {NT state 2 or NT state 3 is satisfied}
        \State Update unplanned airspace zones.
            \State Update air corridors defined by $\mathcal{L}_1$ through $\mathcal{L}_8$.
            \State Update MDP state, transitions, and cost.
        \EndIf
        \If {NT state 1 or NT state 4 is satisfied}
            \State Update MDP cost and transition probabilities.
        \EndIf
        \State {Obtain optimal policy $\pi^*(s)$ ($\forall s\in \mathcal{S}$) using Eq. \eqref{optimalpolicy}.}
    \EndIf
\end{algorithmic}
\end{algorithm}


\section{Simulation Results}\label{sec: Simulation}
In this section, we evaluate the efficacy of the proposed physics-inspired UTM by modeling UAS traffic coordination in the low-altitude airspace above Downtown Tucson. To this end, we use  the data collected from Google-Maps ($x-y$ coordinates and building levels) to generate a 3-D environment with buildings modeled as 3-D objects per Fig. \ref{fig: real map1} and Fig. \ref{fig: 2D map}). Fig. \ref{fig: 2D map floor level}  and  Fig. \ref{fig: 3D matlab} show a top view and $3$-D model of the buildings used by MATLAB to generate the unplanned airspace, defined by $\mathcal{O}_1$ through $\mathcal{O}_{n_o}$.  


\begin{figure}
     \centering
     \begin{subfigure}[b]{0.5\textwidth}
         \centering
         \includegraphics[width=\textwidth]{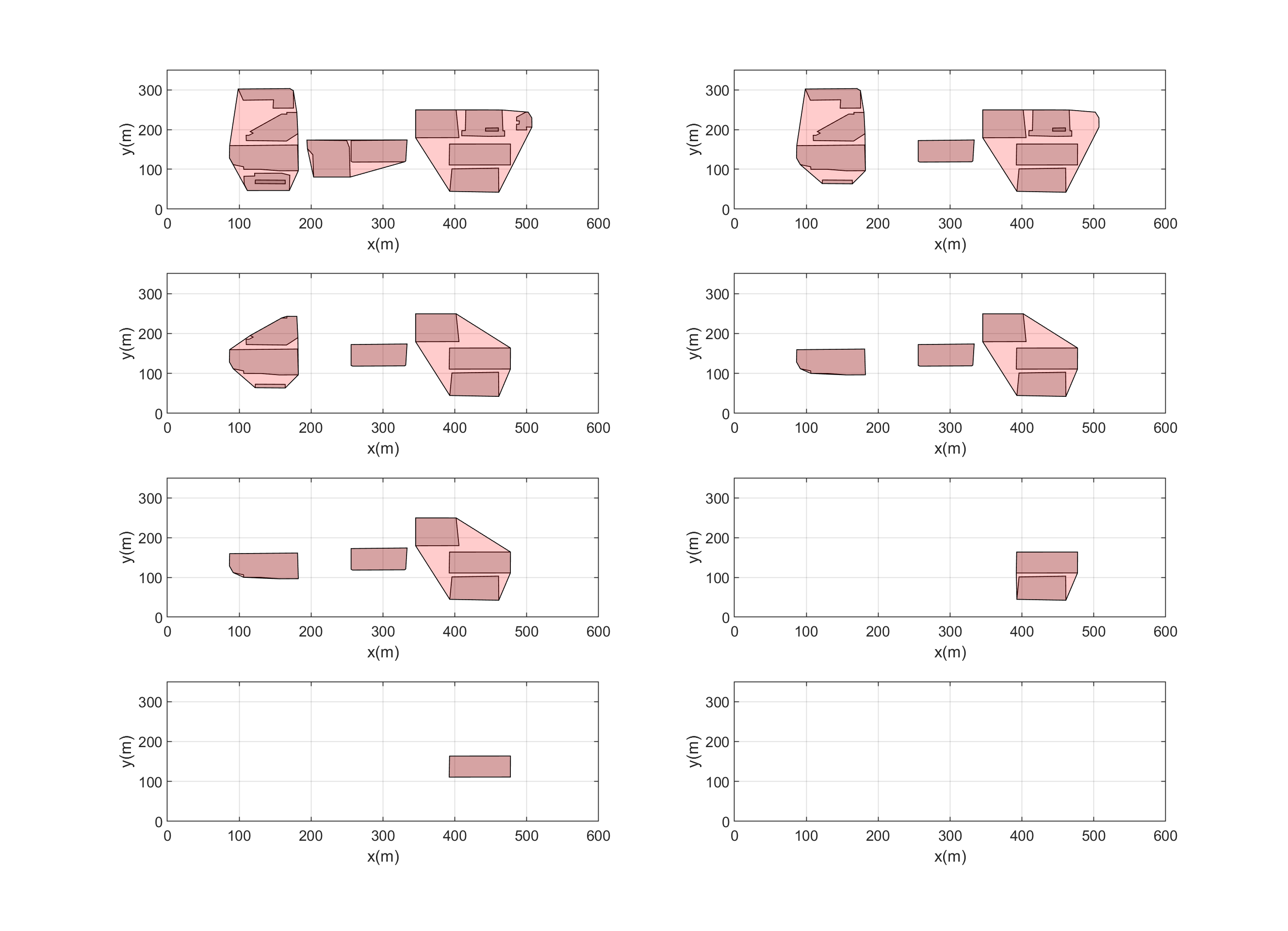}
         \caption{}
         \label{fig: 2D map floor level and obstacles (a)}
     \end{subfigure}
     \hfill
     \begin{subfigure}[b]{0.5\textwidth}
         \centering
         \includegraphics[width=\textwidth]{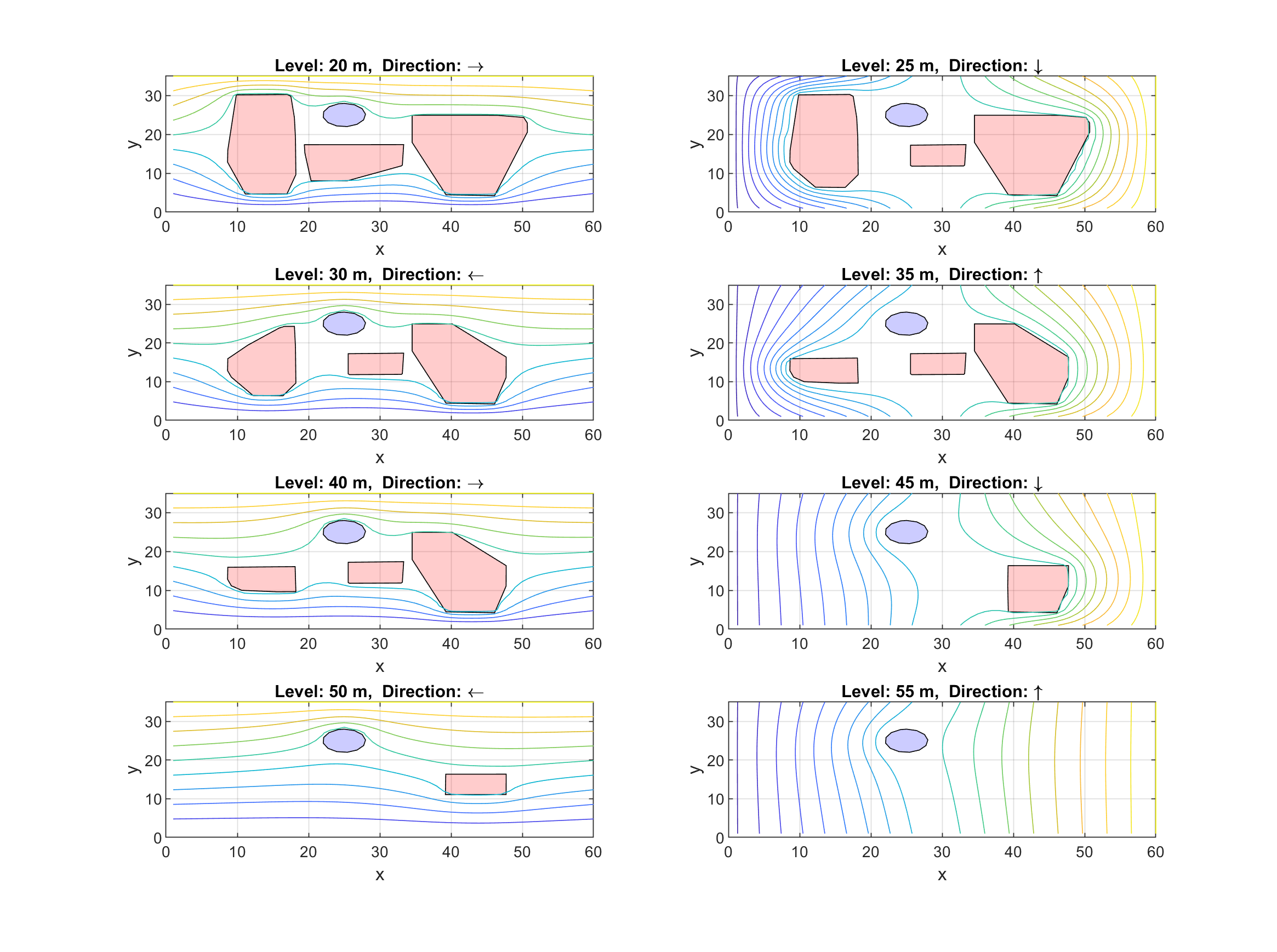}
         \caption{}
         \label{fig: 2D map floor level and obstacles}
     \end{subfigure}
     \hfill
     \begin{subfigure}[b]{0.5\textwidth}
         \centering
         \includegraphics[width=\textwidth]{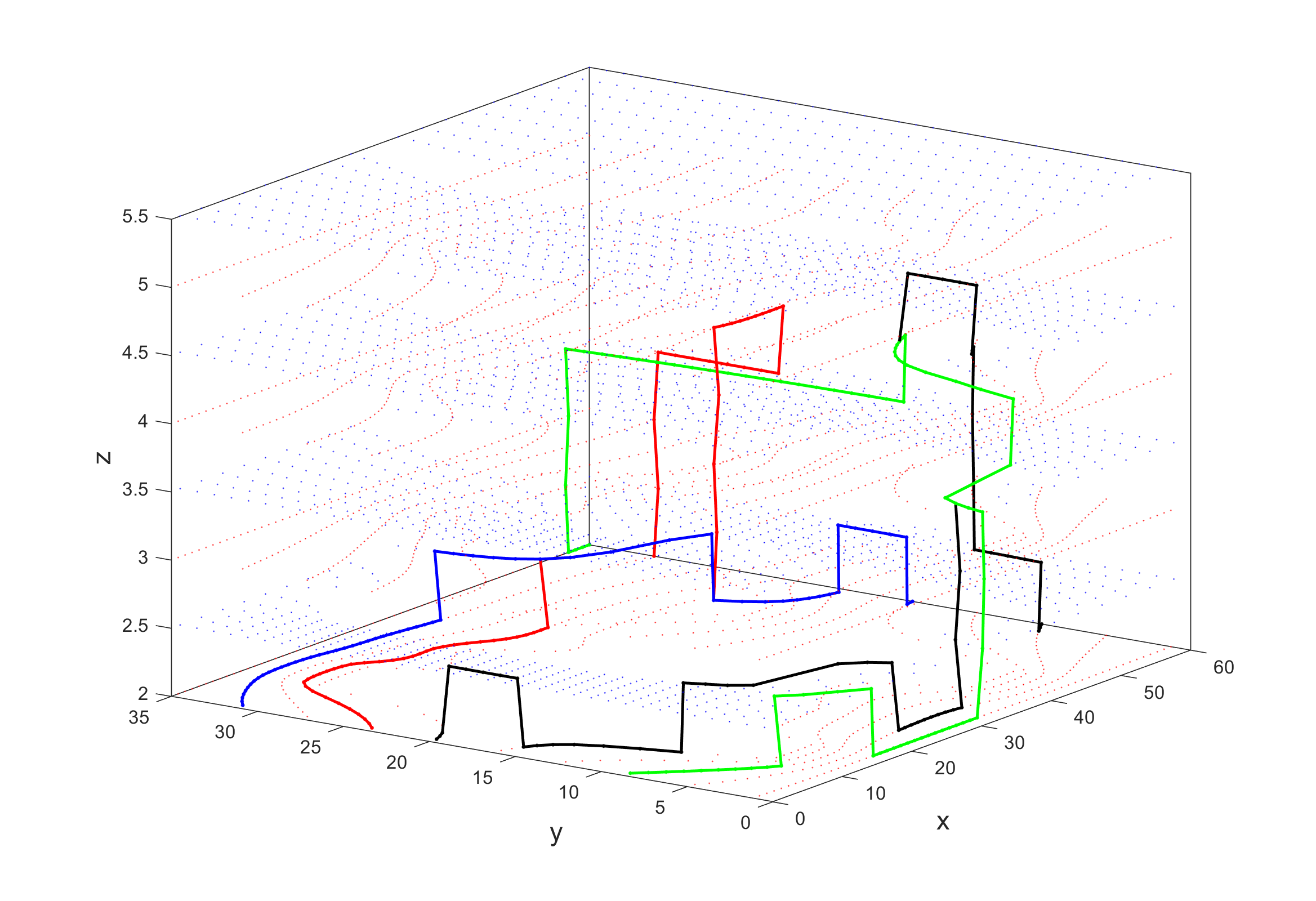}
         \caption{}
         \label{fig: optimal paths}
     \end{subfigure}
        \caption{(a) Cross section of airspace in each layer. Obstacles are shown with colored polygons. (b) Navigable channels in each layer resulted from the ideal fluid flow analysis. (c) Optimal paths obtained from the MDP for four agents with different initial positions and different final destinations.}
        \label{fig: Simulation}
\end{figure}

In order to construct the navigable channels, we divide the airspace, shown in Fig.~\ref{fig: 3D matlab}, into eight layers at altitudes $20m$, $25m$, $30m$, $35m$, $40m$, $45m$, $50m$, and $55m$, respectively. {\color{black}We define $\mathcal{L}_1,\dots,\mathcal{L}_{8}$ as a collection of navigable channels in layers 1 through 8. We define 10 and 18 navigable channels in layers with odd and even index, respectively. A UAS is allowed to move along the corridors with specified direction at every $\mathcal{L}_1$ through $\mathcal{L}_8$.} As shown in Fig.~\ref{fig: 2D map floor level and obstacles}, a UAS is authorized to move in the positive $x$ direction along an air corridor at $\mathcal{L}_1$ and $\mathcal{L}_5$, while it can only move  in the negative $x$ direction inside the channels defined at $\mathcal{L}_3$ and $\mathcal{L}_7$. On the other hand, the corridors in $\mathcal{L}_2$ and $\mathcal{L}_6$ authorize UAS motion along the negative $y$ direction, whereas corridors of $\mathcal{L}_4$ and $\mathcal{L}_8$ permit UAS to move in the positive $y$ direction. In order to simplify the geometrical computational complexity, we use a convex hull operation to combine proximal buildings into a single obstacle (see Fig.~\ref{fig: 2D map floor level and obstacles (a)}). Fig.~\ref{fig: 2D map floor level and obstacles} shows the combined obstacles in different levels. Moreover, we suppose that a manned air vehicle (a helicopter) plans to land on top of the parking building for an emergency case. In this case, ATM defines a no-fly (exclusion) corridor to make a safe flying zone for the helicopter. We model the helicopter's flying zone as a circular cylinder with central axis in the $z$ direction. The blue disks shown in Fig. \ref{fig: 2D map floor level and obstacles} are the projections of the unplanned airspace on $\mathcal{L}_1$ through $\mathcal{L}_8$ allocated by ATM to the helicopter. 
Given the geometry of the unplanned zones, at $\mathcal{L}_1$ through $\mathcal{L}_8$. we generate the navigable channels by using the approach explained in Section~\ref{Time Invariant Navigable Channels}. Fig.~\ref{fig: 2D map floor level and obstacles} shows the streamlines in each layer. We consider 10 and 18 streamlines (corridors) in each layer with motion in $x$ and $y$ directions, respectively.  



We consider a queue of UAS consisting of four UAS requesting transit from departure points $\mathbf{r}_{i,0}$ to destination points  $\mathbf{r}_{i,f}$ in the environment modeled in the previous subsection. In order to define the state set $\mathcal{S}$ we discretize each streamline on each layer. {\color{black}We define $\mathcal{X}$ as grids distributed uniformly every 10$m$ on each streamline in $\mathcal{L}_1,\mathcal{L}_3,\mathcal{L}_5,\mathcal{L}_7$, and similarly, we define $\mathcal{Y}$ as grids distributed uniformly every 10$m$ on each streamline in $\mathcal{L}_2,\mathcal{L}_4,\mathcal{L}_6,\mathcal{L}_8$. Each grid point represents the spatial term of the states in $\mathcal{S}$. Therefore, cardinality of $\mathcal{X},\mathcal{Y}$ is 30 and 10, respectively.}  We assume that UAS enter the airspace in an order of their labelling indices. Implementing Algorithm~\ref{alg: MDP}, we find the optimal policies for each agent. We consider $C_0 = 15$ in the cost function~\eqref{eq: cost function}. Fig.~\ref{fig: optimal paths} shows the optimal paths constructed from Algorithm~\ref{alg: MDP} for a queue of four agents. UAS depart from different points on $x=0$ in the first layer and reach different destination points at $x=60$. Dimensions are scaled by $0.1$ in Fig.~\ref{fig: optimal paths}. Grid points in different layers are shown in Fig.~\ref{fig: optimal paths}.   

     


\section{Conclusion}\label{sec: Conclusion}
This paper has proposed and utilized a novel physics-based method to safely manage low altitude  UAS traffic in low-altitude airspace over a potentially complex urban environment. We used the fundamentals of Eulerian continuum mechanics to spatially define airway corridors around obstacles wrapping buildings and restricted flight zones at low-altitude airspace.  We defined UAS coordination as an ideal fluid flow pattern and obtained geometries of the air corridors by solving Laplace PDEs. For temporal planning, we  define an MDP to model air corridor allocation and managed MDP updates with
a manually-designed finite-state machine.   The efficacy of the proposed method was shown in simulation for low-altitude airspace above Downtown Tucson.


%



\section*{Acknowledgment}
This work has been supported by the National Science Foundation under Award Nos. 2133690 and 1914581.

\ifCLASSOPTIONcaptionsoff
  \newpage
\fi



\bibliographystyle{IEEEtran}
\bibliography{ref}

\begin{IEEEbiography}[{\includegraphics[width=1in,height=1.25in,clip,keepaspectratio]{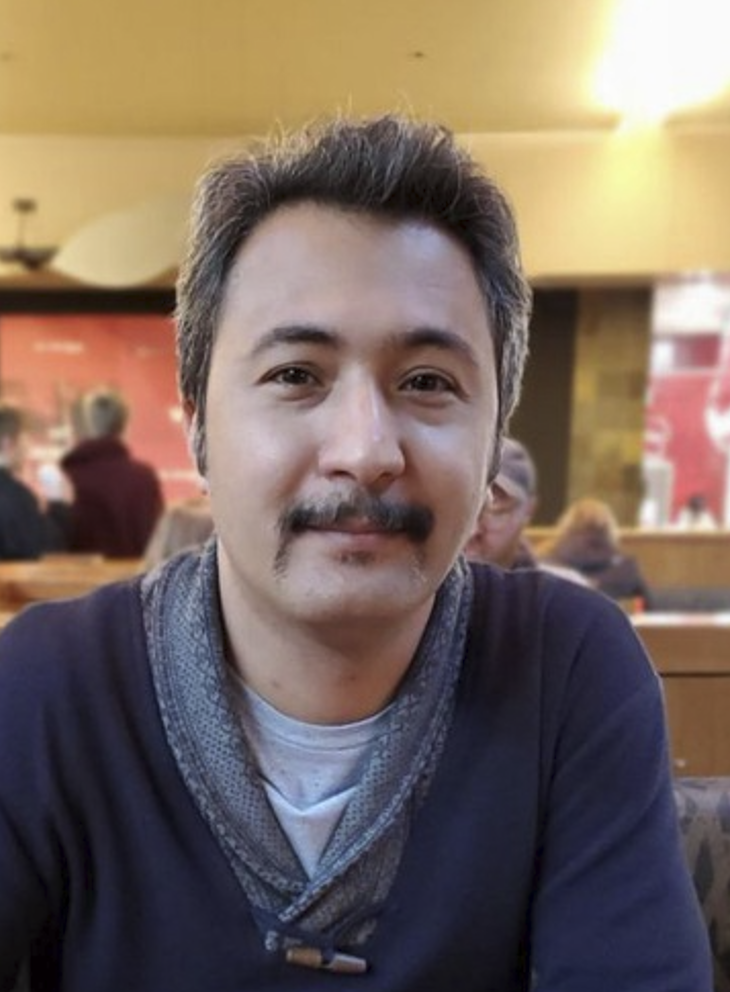}}]
{\textbf{Hamid Emadi}}  received the B.Sc. degree in mechanical engineering from Shiraz University, Shiraz, Iran, the M.S. degree in mechanical systems and solid mechanics from Shiraz University, and the Ph.D. degree in mechanical engineering from Iowa State University, Ames, IA, USA, in 2021. He is a Postdoc research associate in Dr. Rastgoftar's group with the Department of Aerospace and Mechanical Engineering, The University of Arizona, Tucson, AZ, USA.  His research interests include decision making, game theory, optimization, and control.
\end{IEEEbiography}

\begin{IEEEbiography}[{\includegraphics[width=1in,height=1.25in,clip,keepaspectratio]{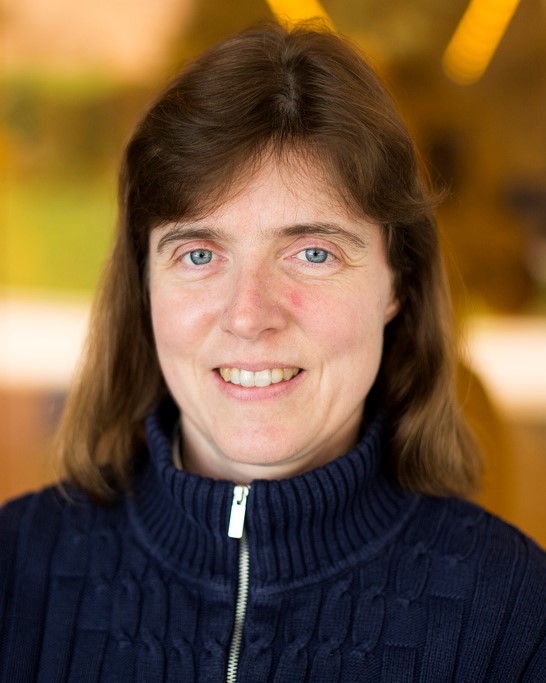}}]
{\textbf{Ella M. Atkins}} is a Professor in the University of Michigan’s Aerospace Engineering Department where she directs the Autonomous Aerospace Systems (A2SYS) Lab and is Associate Director of the Robotics Institute. Dr. Atkins holds B.S. and M.S. degrees in Aeronautics and Astronautics from MIT and M.S. and Ph.D. degrees in Computer Science and Engineering from the University of Michigan. She is an AIAA Fellow, private pilot, and Part 107 UAS pilot. She served on the National Academy’s Aeronautics and Space Engineering Board and the Institute for Defense Analysis Defense Science Studies Group. She has served on several National Academy study committees and co-authored study reports including Advancing Aerial Mobility A National Blueprint (2020) and Autonomy Research for Civil Aviation Toward a New Era of Flight (2014). Dr. Atkins has built a research program in decision-making and control to assure safe contingency management in manned and unmanned Aerospace applications. She is currently Editor-in-Chief of AIAA Journal of Aerospace Information Systems (JAIS) and a member of the 2020-2021 AIAA Aviation Conference Executive Steering Committee.
\end{IEEEbiography}

\begin{IEEEbiography}[{\includegraphics[width=1in,height=1.25in,clip,keepaspectratio]{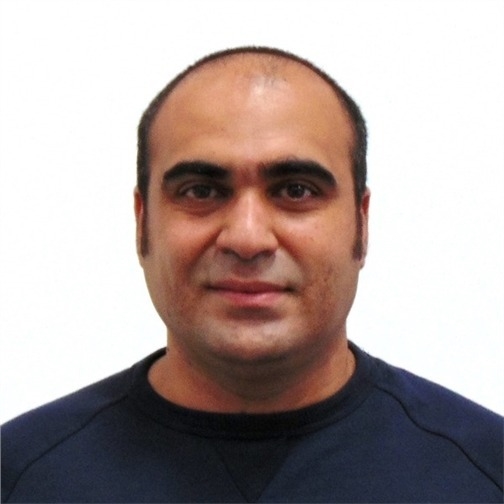}}]
{\textbf{Hossein Rastgoftar}} is an Assistant Professor in the Department of Aerospace and Mechanical Engineering at the University of Arizona and an Adjunct Assistant Professor at the Department of Aerospace Engineering at the University of Michigan Ann Arbor. He received the B.Sc. degree in mechanical engineering-thermo-fluids from Shiraz University, Shiraz, Iran, the M.S. degrees in mechanical systems and solid mechanics from Shiraz University and the University of Central Florida, Orlando, FL, USA, and the Ph.D. degree in mechanical engineering from Drexel University, Philadelphia, in 2015. 
\end{IEEEbiography}
\end{document}